\documentclass[5p,twocolumn,times,number]{elsarticle}

\usepackage{graphicx}
\usepackage{amsmath}   

\begin{document}

\begin{frontmatter}
\title{Towards new Front-End Electronics for the HADES Drift Chamber System}

\author[add1]{M.~Wiebusch\corref{cor}}
\ead{m.wiebusch@gsi.de}
\author[add1]{C.~M\"untz}
\author[add2]{C.~Wendisch}
\author[add2]{J.~Pietraszko}
\author[add1]{J.~Michel}
\author[add1,add2]{J.~Stroth} 
\author[]{for the HADES Collaboration}

\cortext[cor]{Corresponding author}

\address[add1]{Goethe University, Frankfurt, Germany}
\address[add2]{GSI Helmholtzzentrum f\"ur Schwerionenforschung, Darmstadt, Germany}

\begin{abstract}
Operating HADES at the future FAIR SIS-100 accelerator challenges the rate capability of DAQ and
electronics. A new, more robust version of front-end electronics needs to be built for the HADES drift chamber system.
Due to the unavailablitiy of the previously used ASD-8 analog read-out ASIC,
PASTTREC (PANDA straw tube read-out ASIC) was tested as an ASD-8 replacement 
in different scenarios including a beam test.
PASTTREC falls 20\% short of the ASD-8 time precision but performs better w.r.t.~signal charge measurements and overall
operation stability. 
The measured time precision as a function of distance to the sense wire was modeled
within a 3D GARFIELD simulation of the HADES drift cell.

\end{abstract}

\begin{keyword}
Front End, Trigger, DAQ and Data Management - Poster Session

\end{keyword}

\end{frontmatter}

\section{Introduction}
In the HADES set-up 24 Mini (cell) Drift Chambers (MDC) allow for
track reconstruction and determination of charged particle momenta via
deflection in a magnetic field. In addition, MDC supplements
particle identification by measuring the specific energy loss.
Each of the 27000 sensing wires are equipped with a preamplifier, analog pulse shaper and discriminator,
which are combined in the ASD-8~\cite{newcomer} ASIC.
Due to limitations of the current on-board TDCs, especially regarding higher
reaction rates,
the electronics need to be replaced by new
boards featuring multi-hit TDCs. Whereas ASD-8 chips cannot be procured anymore,
a promising replacement candidate is the PASTTREC~\cite{pasttrec} ASIC, developed by JU Krakow.
We tested the ASIC as read-out option for MDC in a variety of set-ups in
direct comparison to ASD-8.
\begin{figure}[h!]
\centering
\includegraphics[width=0.32\textwidth]{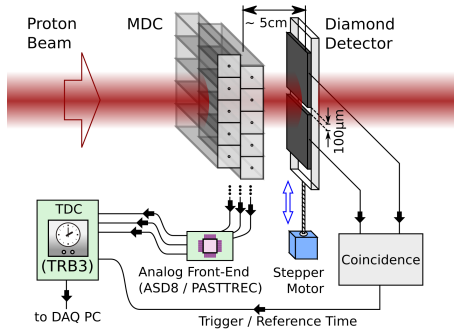}
\caption{Sketch of the COSY beam test set-up to assess timing precision of the joint system comprising a drift chamber and different read-out electronics.}
\label{fig:beam_set-up_simplified}
\end{figure}
\begin{figure}[h!]
\centering
\includegraphics[width=0.35\textwidth]{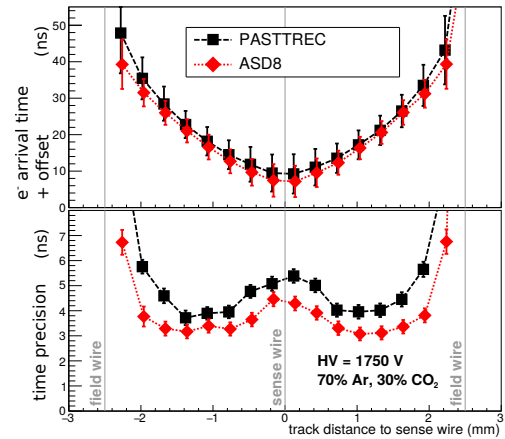}
\caption{Arrival time and time precision of an MDC drift cell read out with ASD-8 and PASTTREC as
a function of track distance to sense wire. 
The arrival time measurement has a systematic offset due to unknown propagation delay in cables and electronics.
The lower plot shows the standard deviation of an asymmetric gaussian fit of the arrival time distribution.}
\label{fig:vw_combo}
\end{figure}

\section{Beam test}
The timing precision
was assessed during a beam test at COSY/J\"ulich using a minimum ionizing proton
beam. As shown in figure \ref{fig:beam_set-up_simplified},
a diamond detector~\cite{diamond} provides reference time and triggers on particles
in a narrow beam slice ($<$100~$\rm{\mu m}$),
aligned with the orientation of the sensing wire, i.e. at near constant distance to it.
The resulting arrival time and arrival time precision is depicted in figure~\ref{fig:vw_combo}.
After applying a walk correction based on the time-above-threshold information,
the PASTTREC falls short of the performance of the ASD-8 by only 20\%.
The difference can be attributed to the longer peaking time of PASTTREC
(15~ns) in contrast to ASD-8 (7~ns).

\section{Laboratory tests}

Apart from testing and preparing the DAQ system for a beam test,
the drift-chamber set-up at the GSI detector lab allowed for complementary measurements.
The two adjacent sense-wire layers
were shifted relative to each other by half the cell width to enable
intrinsic drift time precision measurements
by correllating (summing) the drift times of overlapping cells while
tracking cosmic muons.
The measured precisions are in agreement with the beam test data.

\begin{figure}[b]
\centering
\includegraphics[width=0.33\textwidth]{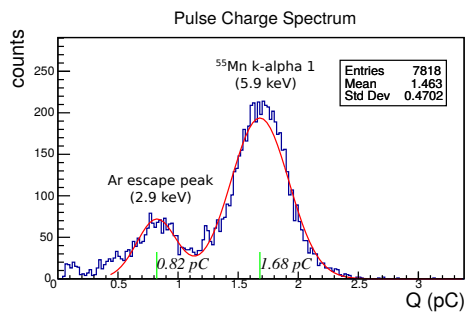}
\caption{Calibrated charge spectrum (derived from the time over threshold distribution) of MDC irradiated with a $^{55}$Fe X-ray source and recorded with PASTTREC.}
\label{fig:1700_Q_hist_peak_desc}
\end{figure}

To study energy loss measurement precision,
a drift chamber was irradiated with an $^{55}$Fe
X-ray source while varying the high voltage.
The pulse charge spectrum (figure~\ref{fig:1700_Q_hist_peak_desc}) is
derived from the time-over-threshold information via a calibration function.
PASTTREC was able to separate the $^{55}$Mn K-alpha peak from the Ar escape peak while varying
the gain by a factor of 15. In the ASD-8 data no clear separation could be observed.

\section{GARFIELD simulation}

\begin{figure}[h!]
\centering
\includegraphics[width=0.28\textwidth]{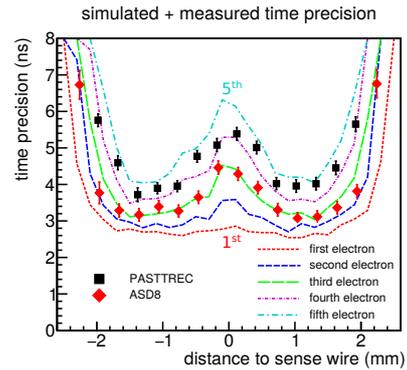}
\caption{Simulated vs measured arrival time precision. Measured precisions plotted with symbols, simulated
precisions drawn with lines. In the simulation the time is measured after an
ideal integrator combined with a discriminator set to integer numbers of electrons arriving at the wire.}
\label{fig:garfield_w}
\end{figure}

The beam test was simulated using 3D GARFIELD~\cite{garfield} and 
the measured drift time as function of track position was reproduced.
To model the characteristic "W" shape of the measured precision,
two additional effects had to be taken into account in the simulation: 
First a gaussian error
added as proxy for the noise at the input of the preamplifier.
Second, the effect of the discriminator threshold  was imitated by waiting for the n-th
fastest electron to arrive at the sensing wire.
As seen in figure~\ref{fig:garfield_w}, the gaussian noise defines the best precision achieved while the integration
over more than one electron deteriorates the resolution in the vicinity of the sensing and potential wires.
According to this model, ASD-8 is sensitive to the third arriving electron, while PASTTREC
reacts to the fourth to fifth arriving electron, as expected due to the different peaking times.

\section{Conclusion}
The PASTTREC ASIC was tested for its suitability to replace
ASD-8 for the read-out of the HADES drift chambers.
While a better time precision is observed when the detector is read out with ASD-8,
a PASTTREC-based read-out seems to benefit the measurement of the deposited energy.
We interpret these findings as an indication that particle identification
in HADES will significantly improve when employing PASTTREC. This benefit might outweigh the loss in spatial precision,
especially because PASTTREC proved to be far less susceptible to pickup noise and self-oscillation than the ASD-8-based FEE,
both in the beam test and the lab test environment.

\section*{Acknowledgments}
This work has been supported by GSI, HIC for FAIR and BMBF (05P15RFFCA).
We would like to thank our colleagues from the group of P.~Salabura at JU Krakow, 
L.~Naumann, X.~Fan from Helmholtz-Zentrum Dresden-Rossendorf and P.~Wintz
from Forschungszentrum J\"ulich.


\end{document}